\documentclass[epj]{svjour}
\usepackage{graphicx}
\usepackage{amssymb}
\usepackage{amsmath}
\usepackage{url}
\usepackage{pythonhighlight}
\usepackage{hyperref}
\begin{document}
\title{SPARKX: A Software Package for Analyzing Relativistic Kinematics in Collision Experiments}
\subtitle{}
\author{Nils Sass\inst{1,2} \and Hendrik Roch\inst{3} \and Niklas G\"otz\inst{1,2} \and Renata Krupczak\inst{4} \and Carl B. Rosenkvist\inst{1,2}
}                     
\offprints{Hendrik Roch}          
\institute{Institute for Theoretical Physics, Goethe University, Max-von-Laue-Strasse 1,
60438 Frankfurt am Main, Germany \and Frankfurt Institute for Advanced Studies,  60438
Frankfurt am Main, Germany \and Department of Physics and Astronomy, Wayne State University, Detroit, Michigan 48201, USA \and Fakult\"at f\"ur Physik, Universit\"at Bielefeld, D-33615 Bielefeld, Germany}
\date{Received: date / Revised version: date}
%
\abstract{
SPARKX is an open-source Python package developed to analyze simulation data from heavy-ion collision experiments. By offering a comprehensive suite of tools, SPARKX simplifies data analysis workflows, supports multiple formats such as OSCAR2013, and integrates seamlessly with SMASH and JETSCAPE/X-SCAPE. This paper describes SPARKX's architecture, features, and applications and demonstrates its effectiveness through detailed examples and performance benchmarks. SPARKX enhances productivity and precision in relativistic kinematics studies.
\PACS{
      {PACS-key}{discribing text of that key}   \and
      {PACS-key}{discribing text of that key}
     } 
} 
\maketitle
\section{Introduction}
Heavy-ion collisions provide a unique opportunity to study nuclear matter under extreme conditions, such as high temperature and density, exploring phenomena like the quark-gluon plasma~\cite{Elfner:2022iae,chaudhuri2014short}. 
To analyze these phenomena, simulations using event generators such as SMASH~\cite{SMASH:2016zqf} and JETSCAPE/X-SCAPE~\cite{Putschke:2019yrg} have become standard in the field. 
However, interpreting the complex datasets generated by these simulations requires sophisticated analysis tools. 
Existing software is often designed with experimentalists in mind, prioritizing performance but frequently resulting in steep learning curves and specialized, inflexible syntax and file formats. 

For theorists and phenomenologists, the lack of user-friendly tools tailored to their specific needs has often necessitated custom-written analysis scripts.
Unfortunately, these scripts are typically untested, prone to errors, and difficult to maintain, posing significant risks to the reliability and reproducibility of the results.
While several well-established analysis frameworks such as ROOT~\cite{ROOT_NIMA_1997} and Rivet~\cite{Bierlich_2020} provide powerful C++-based environments for data analysis, they are primarily designed for experimental workflows and general-purpose data handling rather than streamlined Python-based analysis of event generator outputs.
Consequently, many theory analyses still rely on ad-hoc scripts for reading simulation data, filtering events, and computing observables -- a practice that, without systematic testing, can introduce subtle errors and limit reproducibility.
To address these issues, SPARKX~\cite{hendrik_roch_2025} was developed to bridge this gap, providing a robust, modular, integrated, and rigorously tested Python framework specifically for analyzing event generator outputs from various sources. 
By including pre-implemented observables and adhering to best practices in software design, SPARKX reduces the overhead associated with developing analysis workflows. 
At the same time, its modular and object-oriented design ensures that users can easily extend the codebase to incorporate custom observables and analysis methods. 
This paper introduces SPARKX, describing its motivation, architecture, key features, and practical applications in heavy-ion collision studies. The software architecture of SPARKX follows established principles of modern software engineering, particularly modular and object-oriented design practices as outlined in Martin's Clean Architecture~\cite{martin2}.

The remainder of this paper is organized as follows. 
In Sec.~\ref{sec:program_overview}, we provide an overview of the implemented analysis classes within SPARKX. A quickstart guide to basic functionalities is provided in Sec.~\ref{sec:quickstart}. 
Section~\ref{sec:key_features} describes the key features of the package, highlighting the range of physics analyses that can be performed. 
Implementation details are discussed in Sec.~\ref{sec:implementation_details}, which also introduces the SOLID design principles outlined in Sec.~\ref{sec:design_principles}. 
To benchmark SPARKX against other state-of-the-art analysis tools, we present performance comparisons in Sec.~\ref{sec:preformance_benchmarks}.
The testing routines for verifying the package’s functionality are detailed in Sec.~\ref{sec:testing}, while Sec.~\ref{sec:example_applications} illustrates SPARKX analysis workflows with two example applications. 
Section~\ref{sec:modularity_and_extendability} demonstrates how the framework can be extended to support additional data formats.
Finally, Sec.~\ref{sec:future_development} outlines planned future developments, and Sec.~\ref{sec:conclusion} summarizes our findings and conclusions.

\section{Program Overview} 
\label{sec:program_overview}
SPARKX is designed to provide a comprehensive and modular framework for analyzing heavy-ion collision data in a simple and logical manner. 
The package supports the analysis of outputs from the SMASH and JETSCAPE/X-SCAPE codes. 
Its architecture is class-based, comprising several key components that are introduced in this section with an overview. 

The basic usage of SPARKX is structured around three main steps: first, loading the input file containing particle and event information; second, applying filters to select the relevant data and apply kinematic cuts to a specific problem; and finally, performing the necessary analysis on the selected particles or events. 
Therefore, the input file should contain a particle list from an event generator code, and the produced SPARKX output will contain the final observables from the analysis.

\begin{figure*}
	\centering
	\includegraphics[width=0.9\linewidth]{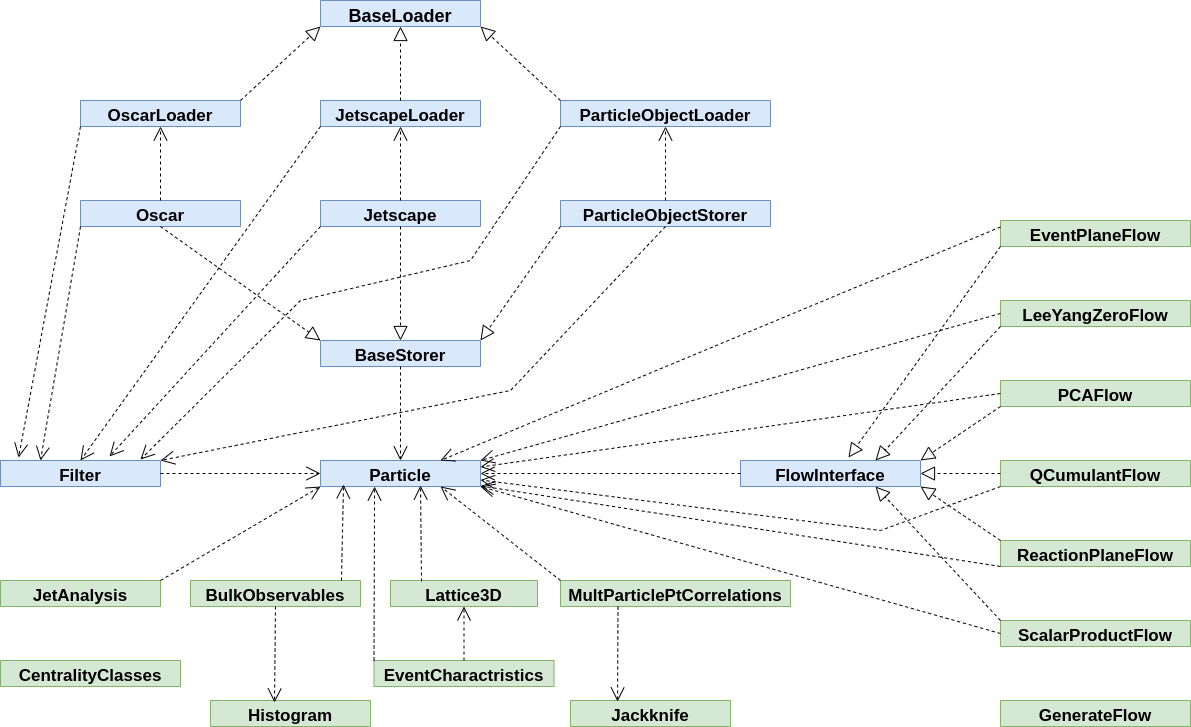}
	\caption{Simplified UML diagram to represent the architecture of SPARKX. Blue boxes represent data storage and processing classes, and green boxes represent more physics quantity-related classes linked to the output of SPARKX. Arrows indicate dependencies between different package parts.}
	\label{fig:diagram}
\end{figure*}

The architecture decisions underlying the code are summarized in the Unified Model Language (UML) diagram in Fig.~\ref{fig:diagram}. 
The blue boxes represent classes responsible for storing and processing the input for the analysis, while the green classes handle the physics calculations. 
The purpose and functionality of each class is elaborated on in the following.
More detailed information about each class can be found in the SPARKX documentation~\cite{documentation}.

\paragraph{Data Loading and Filtering}
The first two steps, loading and filtering, can be done concurrently. The classes \texttt{BaseLoader} and \texttt{Base\-Storer} are an abstract representation of the functionalities, which are implemented specifically for each data format.
\begin{itemize}
    \item \textbf{Storer:}
        These classes store the event data and allow processing and filtering.
        \begin{itemize}
            \item \textbf{\texttt{Base\-Storer}:} Abstract class which defines a ge\-neric object to save event and particle information. All classes below implement it.
            \item \textbf{\texttt{Oscar}:} It handles various OSCAR output formats~\cite{oscar2013,smash_oscar} from different execution modes of SMASH.\footnote{The OSCAR format is a typical output format in hadronic transport codes like SMASH or UrQMD~\cite{Bass:1998ca,Bleicher:1999xi} (implementing an older version of the format).}
            \item \textbf{\texttt{Jetscape}:} Manages hadron or parton outputs from JETSCAPE/X-SCAPE, facilitating seamless data integration. 
            \item \textbf{\texttt{ParticleObjectStorer}:} Saves a list of \texttt{Particle} objects and allows the creation of \texttt{Storer} objects without an external input file.
        \end{itemize}
        \item \textbf{Loader:}
        These classes define how input data is parsed and processed.
        \begin{itemize}
            \item \textbf{\texttt{BaseLoader}:} Abstract class that defines a ge\-neric object to load event and particle information. All classes below implement it.
            \item \textbf{\texttt{OscarLoader}:} Loads and processes different formats of Oscar files.
            \item \textbf{\texttt{JetscapeLoader}:} Loads and processes JET\-SCAPE\\/X-SCAPE files.
            \item \textbf{\texttt{ParticleObjectLoader}:} Overloads the loading me\-thod to process nested lists of \texttt{Particle} objects.
        \end{itemize}
        \item \textbf{Particle Object:}
        \begin{itemize}
            \item \textbf{\texttt{Particle}:} Defines a particle object that serves as the foundation for the SPARKX analysis.
            Additionally, nearly all analysis classes depend on \texttt{Par\-ticle} object instances since these objects are fundamental to all analysis processes.
        \end{itemize}
        \item \textbf{Filtering:}
        \begin{itemize}
            \item \textbf{\texttt{Filter}:} Utility functions that allow filtering of ev\-ents and particles, e.g., to filter only for charged particles or keep only specific kinematic ranges.
        \end{itemize}
\end{itemize}

\paragraph{Analysis}
The final step, after reading and filtering the input file and producing the particle object, is to perform the necessary analysis. All methods assume that prior kinematic cuts and data selection have been applied using filter functionalities. These physics analysis classes are represented by the green boxes in Fig.~\ref{fig:diagram}, with further details on the methods provided in Sec.~\ref{sec:key_features}, since they are more relevant for users. The analyses are categorized into four groups: event-based analysis, flow calculations, jet studies, and a utility group that includes various classes with tools for heavy-ion simulations. 
\begin{itemize}
    \item \textbf{Event Analysis}
        \begin{itemize}
            \item \textbf{\texttt{BulkObservables}:} Computes bulk observables and it can use \texttt{Histogram} as a return type.
            \item \textbf{\texttt{CentralityClasses}:} Determines centrality classes for a set of events.
            \item \textbf{\texttt{EventCharacteristics}:} Evaluates event properties like eccentricities and energy densities, which are crucial for initializing hydrodynamical simulations.
            \item \textbf{\texttt{MultParticlePtCorrelations}:} Calculates multi-particle transverse momentum $p_{\rm T}$ correlations and cumulants. It can be linked to \texttt{Jackknife}~\cite{mcintosh2016jackknifeestimationmethod} for the error calculations.
        \end{itemize}
        \item \textbf{Flow Calculations}: Implements different methods to calculate anisotropic flow (more details in Sec.~\ref{sec:key_features}).
        \begin{itemize}
            \item \textbf{\texttt{EventPlaneFlow}}
            \item \textbf{\texttt{LeeYangZeroFLow}} 
            \item \textbf{\texttt{PCAFlow}}
            \item \textbf{\texttt{QCumulantFlow}} 
            \item \textbf{\texttt{ReactionPlaneFlow}} 
            \item \textbf{\texttt{ScalarProductFlow}}
            \item \textbf{\texttt{FlowInterface}:} It is a class that implements every flow analysis, being more abstract, it is represented in blue color in Fig.~\ref{fig:diagram}.
            \item \textbf{\texttt{GenerateFlow}:} Creates a list of particles with a pre-defined anisotropic flow value for testing purposes. 
        \end{itemize}
        \item \textbf{Jet Analysis:}
        \begin{itemize}
            \item \textbf{\texttt{JetAnalysis}:} Employs the FastJet~\cite{Cacciari:2011ma} package to identify and analyze jets within events. 
        \end{itemize}
        \item \textbf{Utilities:}
        \begin{itemize}
            \item \textbf{\texttt{Histogram}:} Enables the creation and manipulation of histograms for data representation.
            \item \textbf{\texttt{Jackknife}:} Provides methods for statistical error analysis using the Jackknife resampling technique. 
            \item \textbf{\texttt{Lattice3D}:} Manages 3D lattice structures, supporting spatial analyses within the simulation data.
        \end{itemize}
\end{itemize}

This modular design ensures that SPARKX is comprehensive and adaptable, allowing researchers to tailor the toolset to their specific analysis requirements in heavy-ion collision studies. This modular design is also flexible for extensions.

\section{Quickstart}
\label{sec:quickstart}
In this section, we present a brief overview of SPARKX and its application in real-world scenarios. 
For an in-depth exposition of the package's capabilities, please refer to the SPARKX documentation website~\cite{documentation} and the quickstart guide for new users~\cite{documentation_quickstart}. The installation of SPARKX is straightforward via the Python package installer using the command:
\[
\text{\texttt{pip install sparkx}}.
\]
\noindent Once installed, using SPARKX follows a structured workflow:
\begin{enumerate}
    \item Load an input file (either an Oscar or JETSCAPE/X-SCAPE file) by instantiating the appropriate class \texttt{Os\-car} or \texttt{Jetscape}, respectively.
    \item Apply filters and cuts using the corresponding member functions provided by the chosen class (a complete list is available in Ref.~\cite{documentation}). Thereafter, store the refined particle set using \texttt{.particle\_objects\_list()}. This method stores each record as a \texttt{Particle} object, granting access to particle-specific attributes and methods.
    \item Perform further analyses on either the entire particle collection or a selected subset by employing the analysis classes described in Sec.~\ref{sec:program_overview}.
\end{enumerate}
In the following example, we demonstrate the usage of SPARKX with an Oscar file (named \texttt{particle\_lists.os\-car}), which is assumed to exist in the same directory as the script for this case (the procedure for a JETSCAPE/X-SCAPE file is analogous, with the only difference being the use of the \texttt{Jetscape} class). 
In this example, our goal is to generate a differential transverse momentum spectrum and save the resulting data to a file. 
To achieve this, we proceed as follows:
\begin{enumerate}
\item Load the Oscar file.
\item Retain only charged particles and apply the appropriate detector cuts (using exemplary parameter values in this demonstration).
\item Compute and bin $p_\text{T}$ for all remaining particles across all events.
\item Calculate the differential spectrum and the statistical errors.
\item Save the results to a file. 
\end{enumerate}
While this task may seem extensive at first glance, with SPARKX its core implementation is reduced to just six essential lines of code (see Fig.~\ref{fig:quickstart_example}).
\begin{figure*}[!tb]
\begin{python}
from sparkx import Oscar
from sparkx import Histogram

# Define particle filters for the Oscar class
all_filters={'charged_particles': True, 'pT_cut': (0.15, 3.5), 'pseudorapidity_cut': 0.5}
             
# Filter and store data as 2D list with Particle objects
data = Oscar("./particle_lists.oscar", filters=all_filters).particle_objects_list()

# Create a charged hadron transverse momentum list of 
# each particle for all events
pT = [particle.pT_abs() for event in data for particle in event]

# Create a histogram of the transverse momentum with
# 10 bins between 0.15 and 3.5 GeV and add data
hist = Histogram(bin_boundaries=(0.15, 3.5, 10))
hist.add_value(pT)

# Normalize to the number of events and divide by bin width
hist.scale_histogram(1./(len(data)*hist.bin_width()))

# Define the names of the columns in the output file
column_labels = [{'bin_center': 'pT [GeV]',
                'bin_low': 'pT_low [GeV]',
                'bin_high': 'pT_high [GeV]',
                'distribution': '1/N_ev * dN/dpT [1/GeV]',}]
                
# Write the histogram to a file
hist.write_to_file('pT_histogram.dat', hist_labels=column_labels)
\end{python}
\caption{Example code including loading data, applying filters, creating a histogram, and saving a $p_{\rm T}$ spectrum to a file.}
\label{fig:quickstart_example}
\end{figure*}

More examples for the different classes and physics analysis implementations in SPARKX can be found on the documentation website for the individual classes in Ref.~\cite{documentation}.

\section{Key Features}
\label{sec:key_features}
Here, we aim to give an overview of some core features of SPARKX.
\subsection{Data Loading and Filtering}
\label{subsec:load_and_filter}
One of the key features of SPARKX lies in the easy preprocessing of simulation output data.
For example, it offers the possibility to read in simulation output from the SMASH transport code in various formats, e.g., \texttt{Oscar2013} and extensions of it, in one line, and perform filtering operations on particle quantities or for whole collision events.
The same filtering options are offered for JETSCAPE/X-SCAPE hadron or parton output.
SPARKX reads in files event-by-event, creates \texttt{Particle} objects, and then performs the particle and/or event filtering.
Users can implement multiple filtering options in the analysis by providing a dictionary containing all filters for the data.
Here, it is also important to mention that the filters are applied in the order they are included in the dictionary, since the order of filter application is relevant in some cases.
This allows for easy integration into more complex analyses and spares the user from implementing different types of filters themselves.
Implementing these filters to isolate specific event characteristics for the targeted analysis in a well-tested code base (see Sec.~\ref{sec:testing}) improves the reliability of the extracted results, in contrast to an analysis that has these functionalities built in, but is not thoroughly tested.

\subsection{Event Characterization}
\label{subsec:event_characterization}
Another key feature of SPARKX is that it comes with a variety of different classes to perform analyses on loaded (and filtered) simulation outputs (see Sec.~\ref{subsec:load_and_filter}).
We will introduce each of the features here briefly.

In case minimum bias simulations were carried out and one wants to determine the boundaries of centrality classes, this is possible with the \texttt{CentralityClasses} class.
For a given set of event multiplicities, it determines the centrality classes and can, based on this, give the centrality bin of an event in the analysis.

For input data providing position information,~e.g., from SMASH, we provide the \texttt{EventCharacteristics}\\ class to compute spatial event eccentricities in polar coordinates~\cite{Alver:2010gr,Teaney:2010vd,Gardim:2011xv} with the possibility to put different radial weights.
In the case of the first harmonic, the radial weight $r^3$ is used.
The class provides the option to compute the spatial eccentricities on the particle level and also from a smeared density on a grid.
It has the option to use the number, energy, electric charge, baryon number, or strangeness as a weight in the averaging process.
This class also provides the option to smear particles on a 3D grid as number, energy, baryon, electric charge, or strangeness density and write these densities to a file as initial conditions for a hydrodynamic simulation.
The final smearing result can be stored in Milne~\cite{Milne1935} $(\tau,x,y,\eta_s)$ or Minkowski~\cite{Minkowski1908Space} $(t,x,y,z)$ coordinates and the smearing can be performed either using a Gaussian smearing kernel or a Covariant smearing kernel taking the Lorentz contraction of moving particles into account~\cite{Oliinychenko:2015lva}.

The class \texttt{MultiParticlePtCorrelations}, which allows for the computation of multi-particle transverse momentum correlations and cumulants up to the 8-th order, is described in detail in Ref.~\cite{Nielsen:2023znu}.

Another class to compute basic observables from a list of \texttt{Particle} objects is the \texttt{BulkObservables} class.
After loading and filtering, it can compute the event-averaged yields as a function of rapidity, transverse momentum, pseudorapidity, or transverse mass, and return them as SPARKX \texttt{Histogram} objects.
It also allows for the computation of the aforementioned quantities at midrapidity and return their values.

\subsection{Flow Analysis}
SPARKX implements different state-of-the-art methods for anisotropic flow analysis:
\begin{itemize}
    \item \texttt{EventPlaneFlow}~\cite{Poskanzer:1998yz}: The computed anisotropic flow is corrected for the finite multiplicity resolution of the event plane extraction. The event plane resolution is computed by a division of the event into two subsets, one at positive, and one at negative pseudorapidity, with a gap in between.
    \item \texttt{LeeYangZeroFlow}~\cite{Bhalerao:2003yq,Bhalerao:2003xf,Borghini:2004ke}: This method relates the first zero on the real axis of a generating function to the value of the anisotropic flow in the system.
    \item \texttt{PCAFlow}~\cite{Bhalerao:2014mua,Verweij2017}: Differential anisotropic flow from two-particle correlations with Principal Component Analysis.
    \item \texttt{QCumulantFlow}~\cite{Bilandzic:2012wva,Bilandzic:2010jr}: This class implements the 2-nd, 4-th and 6-th order cumulants for anisotropic flow.
    \item \texttt{ReactionPlaneFlow}~\cite{Voloshin:1994mz}: Assumption that the reaction plane is constant for all events,~i.e., the impact parameter is always oriented in the same direction. This method is only suitable for theoretical calculations, not for comparison to experimental data.
    \item \texttt{ScalarProductFlow}~\cite{STAR:2002hbo}: This implementation correlates the event flow vector $Q_n$ with the conjugate unit momentum vector of the particle. It is then normalized by the square root of the scalar product of event flow vectors $Q_n$ from two equally sized sub-events. Therefore, this implementation is only applicable to symmetric collision systems.
\end{itemize}
This variety of different anisotropic flow implementations allows the comparison of the simulation outputs to the results of different experimental collaborations using different techniques in their analyses.
In most of the flow analysis classes, SPARKX provides integrated and differential flow implementations, where the differential flow can be computed as a function of transverse momentum, rapidity, and pseudorapidity.

The \texttt{GenerateFlow} class was introduced to create particle data with a fixed value of anisotropic flow for testing purposes.
Part of this class, where dummy events with a more realistic transverse momentum differential flow are created, is based on an implementation by Nicolas Bor\-ghini~\cite{borghini_generator}.

\subsection{Jet Analysis}
For the analysis of partonic or hadronic jets, SPARKX provides the \texttt{Jet\-Analysis} class, which is a wrapper for the FastJet~\cite{Cacciari:2011ma} Python package.
The wrapper allows setting the jet radius, pseudorapidity range, transverse momentum range, specification of the jet finding algorithm, and whether the associated particles should contain only charged particles.
Outputs of the jet finding algorithm are stored in a separate output file and can be read again with an instance of the \texttt{Jet\-Analysis} class to extract only the jet information or the associated particles in the jet cone for further analysis.

\section{Implementation Details}
\label{sec:implementation_details}
SPARKX follows object-oriented programming. 
The core structure of SPARKX is the \texttt{Particle} class. 
Each particle is represented by an instance of this class. 
This allows for an intuitive implementation of procedures, as particle properties, like rapidity, can be calculated directly from each \texttt{Particle} object on request, and do not depend on meta-structures like events or event lists. 
However, this also poses the challenge of efficient data storage and processing. 
The properties of the particles cannot be stored as attributes of the \texttt{Particle} class, as in Python, this would result in the storage of \texttt{int} and \texttt{double} objects, rather than values of \texttt{int} and \texttt{double}. 
This would result in a greatly increased memory consumption. 
Instead, the \texttt{Particle} class is a wrapper of a NumPy~\cite{numpy} array containing all properties encoded as \texttt{double}. 
This minimizes the overhead while at the same time ensuring full access to all quantities on a particle level. 
To access the property of a particle, the respective entry in its 1D NumPy-array is processed. 
Particle properties are automatically enriched using the latest Particle Data Group (PDG) particle data~\cite{ParticleDataGroup:2024cfk}, with the help of the \texttt{particle} Python package~\cite{eduardo_rodrigues_2025_15543702}.

Data structures like events are represented by nested lists of \texttt{Particle} objects, where a sublist represents a single event. 
These structures are wrapped in their own classes, all inheriting from \texttt{Base\-Storer}. 
These wrappers also store additional data relevant to the whole event, like impact parameters. 
Upon construction, each of these classes calls its respective \texttt{Loader} class, which handles input file parsing. 
This separation ensures efficient handling of dependencies.

Depending on its nature, a \texttt{Loader} object first identifies the file type and checks if it has a valid structure. 
It then loads the data, applies filters, and adds the remaining data to the \texttt{Storer} instance. 
The advantage of this strategy in comparison to a one-pass read-in is twofold: on the one hand, we can validate the structure of the input file and of the event filtering requested by the user before the creation of the \texttt{Particle} objects. 
On the other hand, during the second pass, we can skip directly to the requested events without the overhead of conditionals.

SPARKX relies on several core Python packages to ensure efficient numerical computations and data analysis. 
The main dependencies are:
\begin{itemize}
    \item \texttt{particle}~\cite{eduardo_rodrigues_2025_15543702} for handling particle properties and PDG data,
    \item \texttt{NumPy}~\cite{numpy}, for efficient numerical arrays,
    \item \texttt{SciPy}~\cite{scipy}, for scientific computing routines,
    \item \texttt{FastJet}~\cite{fastjet,Cacciari:2011ma}, for jet finding and clustering algorithms,
    \item \texttt{Matplotlib}~\cite{matplotlib}, for plotting and visualization of results.
\end{itemize}

\section{Design Principles}
\label{sec:design_principles}
SPARKX follows the SOLID design principles of object-oriented programming \cite{martin2000design}. 
In practice, this is enforced in the following way:
\begin{itemize}
    \item \textbf{Single-responsibility principle}: Every class has only one responsibility. This is the core reason for separating data storage and loading, as well as input file processing and filtering into different classes. This principle ensures the isolation of different regions of the codebase and well-defined responsibilities, ensuring high maintainability.
    \item \textbf{Open-closed principle}: This principle states that software entities shall be open for extension, but closed for modification. We enforce this, for example, in the flow classes, which all inherit from an abstract base class. The abstract base class defines the interface, which has to remain unchanged. Users can trust that this interface will not change in future versions, allowing them to use SPARKX as a library for long-term projects. On the other hand, extensions are encouraged by enriching the \texttt{Flow} module with further classes that implement different algorithms using the same base class.
    \item \textbf{Liskov substitution principle}: The abstractions applied in SPARKX follow the Liskov substitution principle, which in principle states that classes that inherit from parent classes should be able to substitute them without breaking the program. The strict adherence to this allows the polymorphic behavior -- for practically any operation with SPARKX, it is irrelevant whether the user derived their data from a JETSCAPE/X-SCAPE or a SMASH simulation, as the \texttt{Storer} classes both fulfill the same design contract.
    \item \textbf{Interface segregation principle}: According to this principle, clients should not be forced to depend upon interfaces that they do not use. In our case, we restrict the shared interfaces of all storing and loading classes to a minimum, which decouples the classes and enforces a high degree of modularity.
    \item \textbf{Dependency inversion principle}: This principle\\ tries to avoid the traditional layer system, where low-level layers are consumed by high-level layers. This has the downside that high-level components depend on many low-level implementations, hindering maintainability and flexibility due to coupling.  Instead, one uses interfaces that the high-level module depends on and the low-level module implements. We realize this, for example, in the relationship between \texttt{Flow} and storage classes. \texttt{Flow} depends only on the \texttt{Base\-Storer} abstract class and is thus completely independent of changes in the low-level classes.
\end{itemize}
With the application of these design principles, we ensure high maintainability and simplify further extensions. 
Indeed, implementing further features has a reduced risk of breaking existing code. 
Due to the high encapsulation and clear architecture, changes are highly localized, which reduces the cost of modification further.

\section{Performance Benchmarks}
\label{sec:preformance_benchmarks}
An important consideration for any analysis, especially in heavy-ion collisions, is the memory usage and execution time of the analysis program. 
SPARKX is a modern Python-based analysis framework designed for flexibility and ease of use. To benchmark its performance, we compare it to Rivet~\cite{Bierlich_2020}, a widely used C++ framework for analysis algorithms. 
As a test case, we implement a simple transverse momentum analysis for charged particles in Pb-Pb collisions at $\sqrt{s_{\rm NN}} = 17.3\;\mathrm{GeV}$ using both SPARKX and Rivet.

\begin{figure*}[!tb]
    \centering
    \begin{minipage}{0.5\linewidth}
        \centering
        \includegraphics[width=\linewidth]{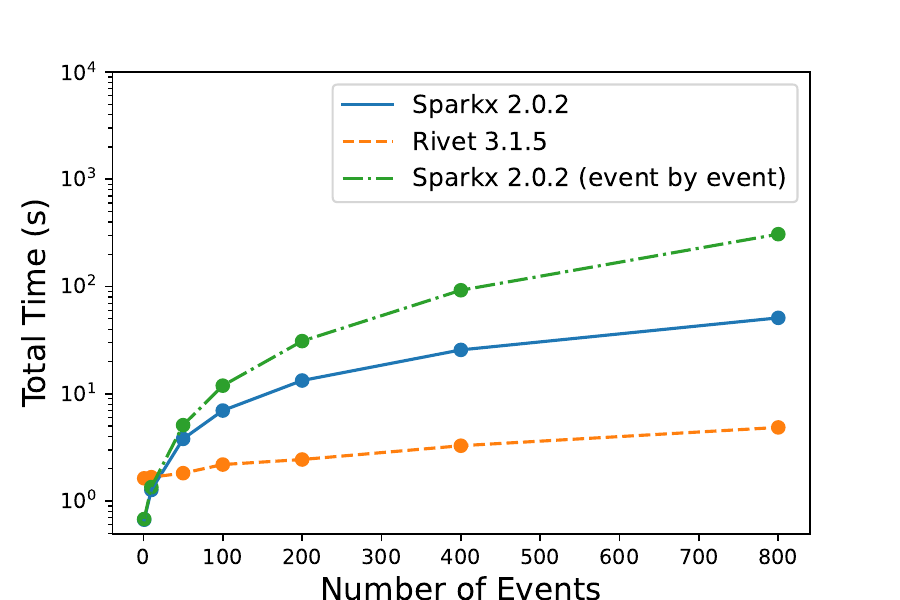}
    \end{minipage}\hfill
    \begin{minipage}{0.5\linewidth}
        \centering
        \includegraphics[width=\linewidth]{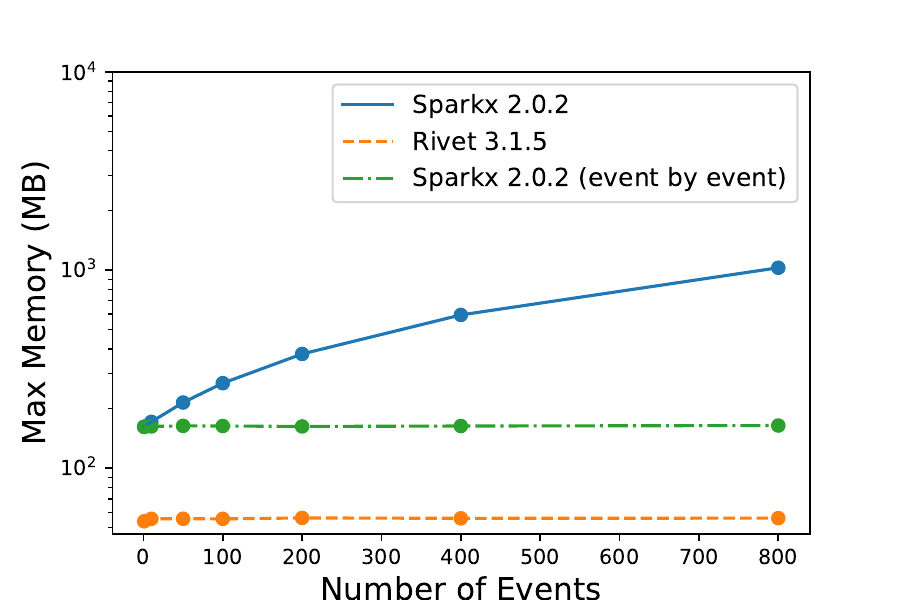} 
    \end{minipage}
    \caption{Execution time (left) and maximum memory usage (right) for a simple charged hadron transverse momentum analysis implemented in SPARKX 2.0.2 and Rivet 3.1.5.}
    \label{fig:both_benchmarks}
\end{figure*}
As shown in Fig.~\ref{fig:both_benchmarks}, SPARKX has a longer execution time than Rivet, primarily due to two factors. First, SPARKX is implemented in Python, which prioritizes usability over raw performance, whereas Rivet is written in C++. 
Second, SPARKX, by default, loads the entire Oscar file into memory, enabling efficient analysis of large datasets, but at the cost of higher memory usage. 
In contrast, Rivet processes events sequentially, maintaining a nearly constant memory footprint.
SPARKX also offers an event-by-event reading mode, which significantly reduces memory consumption but increases execution time, as shown in the panels of Fig.~\ref{fig:both_benchmarks} by the dash-dotted lines.
However, this overhead can be mitigated through more efficient event parsing. 
In most analyses, data can be read event by event unless full event statistics, such as correlation functions, are required. 
Consequently, the overall memory usage is typically not a concern, especially when filters are applied during reading, since they restrict the data loaded into memory. 
This underscores that the package is designed to balance ease of use and resource efficiency rather than to maximize performance.

Future development will focus on optimizing data loading strategies to improve both execution time and memory efficiency. 
Planned enhancements include more efficient Oscar file parsing and improved parallel processing to narrow the performance gap while preserving the flexibility of a Python-based framework.

\section{Testing}
\label{sec:testing}
A core aim of SPARKX is not just to provide rapid analysis but also to ensure a high degree of stability and reliability in the results. 
To achieve this, we have adopted a comprehensive testing strategy that includes various levels of testing, ensuring robustness and correctness across all stages of development and deployment.
While providing a wide range of pre-implemented observables, we employ meticulous testing to ensure correct results consistently, also between releases. 
The backbone of our tests is unit tests for quality assurance, which check individual functions and methods for correctness. 
Unit tests are employed extensively to validate specific components and functions in isolation, catching issues early in the development cycle. 
Integration tests, on the other hand, guarantee that different modules of SPARKX operate seamlessly together. 
These tests specifically verify the interfaces between different components, ensuring that the integrated system maintains expected behavior as changes are introduced. 
A typical example for these tests is integration tests for the \texttt{EventCharacteristics}, which not only check individual methods, but also the interface between \texttt{Lattice3D} and  \texttt{EventCharacteristics}, as the former is used in the latter.
Beyond just test coverage, which currently stands at 90\%, our approach emphasizes strategically designing tests that address the most critical and susceptible components of the system, thus maintaining a consistently high level of reliability even as the codebase evolves. 
We have focused on the creation of tests based on \texttt{pytest}~\cite{pytest} on crucial elements of the code and features that are especially susceptible to changes in future iterations.

Another feature to ensure consistently high code quality is the enforcement of static typing. 
Although Python is dynamically typed, this potentially allows for unexpected behavior when unexpected formats are present in input data. 
By enforcing static typing using the \texttt{mypy}~\cite{mypy} linter, we can ensure that parts of the code that would give ambiguous behavior are identified and eliminated.

The SPARKX GitHub repository makes use of automated workflows to ensure the tests pass and the static typing is fulfilled before pull requests can be merged to the development or main repository branches.
There is also an automated code formatting action using the \texttt{black} Python package~\cite{black}.
This ensures that in each development step, the whole codebase adheres to pre-defined formatting standards and stays functional at all times. 

\section{Example Applications}
\label{sec:example_applications}
We illustrate SPARKX's capabilities through two examples:
\begin{itemize}
    \item \textbf{Flow Analysis (Fig.~\ref{fig:flow_example}):} Using SMASH-generated datasets, SPARKX computes anisotropic flow coefficients of char\-ged particles as a function of collision centrality. We first define an \texttt{Oscar} object to store the events and calculate centrality bins. Then, we exemplify how to calculate the flow for midcentral events.
    \item \textbf{Jet Analysis (Fig.~\ref{fig:jet_example}):} Analysis of hadronic jets from JET\-SCAPE/X-SCAPE output, highlighting the ability of SPARKX to be an efficient wrapper for the FastJet Python package.
    This example loads simulation output from JETSCAPE/X-SCAPE as a list of \texttt{Particle} objects and filters out all neutral particles and particles below 0.1 GeV transverse momentum. Then, it performs a jet finding analysis using the FastJet package and writes the extracted jets with their associated hadrons to a file.
    This file is then read in again, and the jets/associated particles are extracted for further analysis.
\end{itemize}
The SPARKX framework has already been employed in several (published) physics studies, with applications including the definition of centrality classes, the calculation of particle spectra and eccentricities, as well as the analysis of anisotropic flow and jet observables~\cite{Gotz:2023kkm,Gotz:2025wnv,Borghini:2024kll,Andronic:2025ylc,Roch:2025tbx,Roch:2025pcj}.

\begin{figure*}[!tb]
\begin{python}
from sparkx import Oscar, CentralityClasses
from sparkx.flow import QCumulantFlow
from copy import deepcopy
OSCAR_FILE_PATH = "[Oscar_dir]/particle_lists.oscar"

oscar = Oscar(OSCAR_FILE_PATH, filters={'charged_particles':True})

centrality_bins = [0, 20, 40, 60, 100]
events_multiplicity = [event[1] for event in oscar.num_output_per_event_]
centrality_obj = CentralityClasses(events_multiplicity=events_multiplicity, centrality_bins=centrality_bins)
# Generate storers for two centrality classes
upper_bounds = centrality_obj.dNchdetaMax_
lower_bounds = centrality_obj.dNchdetaMin_
oscar_central = deepcopy(oscar).multiplicity_cut(
                    (int(upper_bounds[0]),int(lower_bounds[0])))
oscar_midcentral=oscar.multiplicity_cut((int(upper_bounds[1]),int(lower_bounds[1])))
# calculate flow
flow = QCumulantFlow(n=2, k=2, imaginary='zero')
v2_central, v2_err_central = flow.integrated_flow(oscar_central.particle_objects_list())
v2_midcentral, v2_err_midcentral = flow.integrated_flow(oscar_midcentral.particle_objects_list())
\end{python}
    \caption{Example code to create centrality classes and compute the integrated anisotropic flow $v_2$ for charged hadrons.}
    \label{fig:flow_example}
\end{figure*}

\begin{figure*}[!tb]
\begin{python}
from sparkx import Jetscape
from sparkx.jet import JetAnalysis

JETSCAPE_FILE_PATH = " [Jetscape_directory]/final_state_hadrons.dat"
JET_ANALYSIS_OUTPUT_PATH = "[Jetscape_directory]/jet_analysis_output.dat"

cuts = {'charged_particles': True, 'pT_cut': (0.1, None)}

hadrons = Jetscape(JETSCAPE_FILE_PATH, filters=cuts).particle_objects_list()
# Perform the jet analysis
jet_analysis = JetAnalysis()
jet_analysis.perform_jet_finding(
                hadrons,
                jet_R=0.6,
                jet_eta_range=(-2., 2.),
                jet_pT_range=(10., None), 
                output_filename=JET_ANALYSIS_OUTPUT_PATH)
# Read the jets from file
jet_analysis.read_jet_data(JET_ANALYSIS_OUTPUT_PATH)
jets = jet_analysis.get_jets()
# List of the associated particles for all jets 
# (associated hadrons for each jet have a sub-list)
assoc_hadrons = jet_analysis.get_associated_particles()
# Further analysis can be done with the jets and associated particles
\end{python}
    \caption{Example code to run a jet finding algorithm on the input data and write the jets with associated particles to a file.}
    \label{fig:jet_example}
\end{figure*}
    
\section{Modularity and Extendability}
\label{sec:modularity_and_extendability}
SPARKX is specifically designed to read SMASH and JET\-SCAPE/X-SCAPE formats but is built with flexibility in mind to support additional formats. 
The current implementation uses modular \texttt{Loader} and \texttt{Storer} classes that inherit from the abstract base classes \texttt{BaseLoader} and \texttt{Base\-Storer}. 
With minimal effort, any custom format can leverage these base classes to gain access to essential features such as standardized data structures, diverse filter choices, and compatibility with SPARKX's core analysis workflows. 
Key features enabled by using the base classes include efficient data parsing and filtering for large datasets, seamless integration with SPARKX's analysis modules, and compatibility with SPARKX's particle and event-level operations.

For a fast and easy approach to integrating user data into SPARKX's anisotropic flow and jet analyses, one can utilize the getter functions of the \texttt{Particle} class.
These functions allow formatting the data as a list of \texttt{Particle} objects, which can then be passed to the desired analysis interface.

\subsection{Custom Loader}
The \texttt{BaseLoader} class provides a common interface for all loader classes responsible for loading data from a file.
To implement a custom loader class, the \texttt{BaseLoader} requires the \texttt{\_\_init\_\_} and \texttt{load} methods to be implemented (see Fig.~\ref{fig:custom_loader}). 
\begin{figure*}
\begin{python}
@abstractmethod
def __init__(self, path: str) -> None:
    """
    Abstract constructor method.

    Parameters
    ----------
    path : str
        The path to the file to be loaded.
    """
    pass

@abstractmethod
def load(self, **kwargs: Dict[str, Any]) -> Any:
    """
    Abstract method for loading data.

    Raises
    ------
    NotImplementedError
        If this method is not overridden in a concrete subclass.
    """
    raise NotImplementedError("load method is not implemented")
\end{python}
    \caption{Functions needed to implement a custom loader class.}
    \label{fig:custom_loader}
\end{figure*}

They represent the constructor, which for loaders, main\-ly sets information about the file type, and the actual loading procedure, which generates lists of \texttt{Particle} objects and the event structure.
Specifically, the \texttt{load} method should return the objects shown in Fig.~\ref{fig:load_return}.
\begin{figure}
\begin{python}
self.create_loader(path)
if self.loader_ is not None:
    (
        self.particle_list_,
        self.num_events_,
        self.num_output_per_event_,
        self.custom_attr_list,
    ) = self.loader_.load(**kwargs)
else:
    raise ValueError("Loader has not been 
    created properly.")
\end{python}
    \caption{Required return objects from the \texttt{load} method.}
    \label{fig:load_return}
\end{figure}
In this code example, the \texttt{create\_loader}  method is called with a file path.
The loader's \texttt{load} method is then expected to return a tuple with the following elements:
\begin{itemize}
    \item \texttt{self.particle\_list\_}: The list of \texttt{Particle} objects in sub-lists for each event.
    \item \texttt{self.num\_events\_}: The number of events contained in the list.
    \item \texttt{self.num\_output\_per\_event\_}: The number of particles per event.
    \item \texttt{self.custom\_attr\_list}: A list of custom attributes.
\end{itemize}

\subsection{Custom Storer}
The \texttt{Base\-Storer} class is a generic class designed as a common interface to store loaded particle objects from a file, as well as perform filtering and merging operations.
It contains three abstract methods, \texttt{\_update\_after\_merge}, \texttt{create\_loader}, and \texttt{\_particle\_as\_list}, shown in Fig.~\ref{fig:custom_storer}.
\begin{figure*}
\begin{python}
@abstractmethod
def _update_after_merge(self, other: "Base\-Storer") -> None:
    raise NotImplementedError("This method is not implemented yet")

@abstractmethod
def create_loader(self, arg: Union[str, List[List["Particle"]]]) -> None:
    raise NotImplementedError("This method is not implemented yet")

@abstractmethod
def _particle_as_list(self, particle: "Particle") -> List:
    raise NotImplementedError("This method is not implemented yet")
\end{python}
    \caption{Functions needed to implement a custom storer class.}
    \label{fig:custom_storer}
\end{figure*}
In the initialization of the class, the \texttt{create\_loader} me\-thod can accept either a string (representing a file path) or a nested list of \texttt{Particle} objects. 
This allows both reading event data from a file or directly from an interfacing software. 
However, the loader’s \texttt{load} method must return the required fields mentioned above. 
Apart from this, it is only necessary to define what validation steps have to be performed after merging multiple store objects, as well as how to store contained \texttt{Particle} object lists to lists of primitive data types.

This shows that SPARKX can, in general, support a wide range of file formats and is easily extendable to analyze simulation output from other collider physics simulation packages.
While more efficient formats such as HDF5 or ROOT are not currently supported, this can be considered for future extensions if needed.

\section{Future Developments}
\label{sec:future_development}

Future improvements of SPARKX will focus on extending its data handling, boosting its computational efficiency, and enhancing its core analysis toolkit. 
These developments include adding binary file input and output capabilities so that researchers can process large volumes of collision data without the overhead often associated with text-based formats. 
The ability to read particle interaction blocks in SMASH-generated files will allow more comprehensive reconstructions of particle histories, thereby facilitating in-depth collision event analysis. 
Moreover, an event-by-event read and analysis mode is planned to conserve system memory and make the processing of extensive datasets more feasible. 
Enhancements to parallelization will further optimize data ingestion by distributing multiple events across available CPUs, while certain analysis routines will likewise be parallelized to reduce overall computation time. Substantial performance gains are also to be expected when moving core operations of the codebase which are less exposed to users from Python to C++ via bindings.
Additionally, file reading will be streamlined to reduce the number of passes necessary when accessing large datasets, thereby improving overall performance. 
SPARKX also aims to provide native support for SMASH ensemble-mode files, allowing to extend the analysis also on the simulations including nuclear potentials. 
To broaden the range of correlation studies, the platform will incorporate capabilities for calculating Hanbury Brown-Twiss (HBT) radii. 
Another goal is to introduce a preimplemented functionality that divides a \texttt{Base\-Storer} object into centrality classes, enabling researchers to categorize collisions more comfortably in the analysis workflow.  

SPARKX is an open-source project, and we welcome external contributions, encouraging users to merge their extensions and improvements back into the main repository. 
SPARKX is a software project that not only aims to serve the community but can also be improved and maintained by any interested member of the community.

\section{Conclusion}
\label{sec:conclusion}
SPARKX delivers a unified and modular approach to the analysis of heavy-ion collision data, bridging the gap between raw simulation output and high-level observables. 
Its object-oriented architecture, designed around well-esta\-blished principles, ensures clarity, maintainability, and ease of extension, allowing new analysis features or third-party data formats to be integrated with minimal effort. 
By offering an extensive suite of pre-validated routines -- ranging from bulk observables and anisotropic flow calculations to jet analyses -- SPARKX enables both beginners and experienced researchers to streamline their workflows while minimizing the risks of coding errors. 
Thorough testing, enforced static typing, and clear documentation further contribute to the reliability and transparency of the software.

Currently, SPARKX is primarily used by its developers, their working groups, and students. 
We have also introduced the package to a broader audience during the JETSCAPE Summer School 2023, where it was employed in the ``Hadronization'' exercise session~\cite{JETSCAPE_BrickAnalysis,JETSCAPE_ElectronPositronAnalysis}. 
We hope that the package will gain wider adoption in the community over time.

Looking ahead, SPARKX's development will continue focusing on broadening its scope to handle additional file formats, accelerate performance through refined parallelization strategies and using Python C++ bindings, and expand its physics toolkit with new correlation and event-structure analyses. 
With these future enhancements,\\ SPARKX stands to become an indispensable resource for the community, offering reproducible and efficient workflows that adapt to evolving research needs. 
Ultimately, by alleviating technical barriers and fostering best practices, SPARKX aims to maximize the scientific return of high-energy nuclear collision studies, empowering researchers to delve deeper into emergent phenomena in QCD and beyond.

\begin{acknowledgement}
We sincerely thank Hannah Elfner for her support during the early stages of the SPARKX project and for facilitating the hosting of the project's repository under the \texttt{smash-transport} organization on GitHub, ensuring its long-term sustainability. We are also very grateful to Alessandro Sciarra for the valuable input, inspiration and assistance during the development of this software. We also extend our gratitude to Nicolas Borghini for insightful discussions on the implementation of anisotropic flow algorithms.
H.~R. was supported by the National Science Foundation (NSF) within the framework of the JETSCAPE collaboration (OAC-2004571) and by the DOE (DE-SC0024232). 
This work was supported by the Deutsche Forschungsgemeinschaft (DFG, German Research Foundation) – Project number 315477589 – TRR 211. 
N.~G. acknowledges support by the Stiftung Polytechnische Gesellschaft Frankfurt am Main as well as the Studienstiftung des Deutschen Volkes.
\end{acknowledgement}

\bibliographystyle{ieeetr}
\bibliography{sparkxbib.bib}
\end{document}